\documentclass{pasa}%

\usepackage{graphicx}

\def\aj{AJ}%
\def\actaa{Acta Astron.}%
\def\apj{ApJ}%
\def\apjl{ApJ}%
\def\aap{A\&A}%
\def\aapr{A\&A~Rev.}%
\def\mnras{MNRAS}%
\def\pasp{PASP}%
\def\na{New A}%

\title[New NIR eclipsing binary system candidates]{A hundred new eclipsing binary system candidates studied in a near-infrared window in the VVV survey \thanks{Based on observations taken within the ESO VISTA Public Survey VVV, Programme ID 179.B-2002.}}

\author[L. V. Gramajo et al.]{
L.V. Gramajo,$^{1,2}$\thanks{E-mail: luciana@oac.unc.edu.ar}
T. Palma,$^{1,2}$
D. Minniti,$^{3,4}$
R. K. Saito,$^{5}$
J.J. Clari\'a,$^{1,2}$
R. Kammers,$^{5}$
F. Surot$^{6,7}$
\affil{$^{1}$Universidad Nacional de C\'ordoba. Observatorio Astron\'omico de C\'ordoba, C\'ordoba, Argentina}
\affil{$^{2}$Consejo Nacional de Investigaciones Cient\'ificas y T\'ecnicas (CONICET), Godoy Cruz 2290, Buenos Aires, CPC 1425FQB, Argentina}
\affil{$^{3}$Departamento de Ciencias F\'isicas, Facultad de Ciencias Exactas, Universidad Andr\'es Bello, Av. Fernandez Concha 700, Las Condes,\\ Santiago, Chile}
\affil{$^{4}$Instituto Milenio de Astrof\'isica, Santiago, Chile}
\affil{$^{5}$Departamento de F\'{i}sica, Universidade Federal de Santa Catarina, Trindade 88040-900, Florian\'opolis, SC, Brazil }
\affil{$^{6}$European Southern Observatory, Karl Schwarzschild-Stra$\beta$e D-85748, Garching bei M\"unchen, Germany }
\affil{$^{7}$Instituto de Astrof\'isica, Pontificia Universidad Cat\'olica de Chile, Av. Vicu\~na Mackenna 4860, Santiago , Chile.}}

\jid{PASA}
\doi{10.1017/pas.\the\year.xxx}
\jyear{\the\year}

\usepackage{aas_macros}
\usepackage{hyperref} 
\hypersetup{colorlinks,citecolor=blue,linkcolor=blue,urlcolor=blue}

\hypersetup{draft}

\begin{document}

\begin{frontmatter}
\maketitle

\begin{abstract}

We present the first results obtained from an extensive study of eclipsing binary (EB) system candidates recently detected in the VISTA Variables in the V\'ia L\'actea (VVV) near-infrated (NIR) Survey. We analyze the VVV tile d040 in the southern part of the Galactic disc wherein the interstellar reddening is comparatively low, which makes it possible to detect hundreds of new eclipsing binary candidates. We present here the light curves and the determination of the geometric and physical parameters of the best candidates found in this ``NIR window'', including 37 contact, 50 detached and 13 semi-detached eclipsing binary systems. We infer that the studied systems have an average of the $K_s$ amplitudes of $0.8$ mag and a median period of 1.22 days where, in general, contact binaries have shorter periods.  Using the ``Physics Of Eclipsing Binaries'' (PHOEBE) interactive interface, which is based on the Wilson and Devinney code, we find that the studied systems have low eccentricities.
The studied EBs present mean values of about 5700 K and 4900 K for the $T_1$ and $T_2$ components, respectively. The mean mass-ratio ($q$) for the contact EB stars is $\sim$ 0.44. This new Galactic disc sample is a first approach to the massive study of NIR EB systems. 

\end{abstract}

\end{frontmatter}

\section{Introduction} \label{sec:intro}

About 70 per cent of the stars in our Galaxy are believed to be part of binary or multiple stellar systems \citep{duque91}. Such systems are excellent laboratories not only to examine the physical properties of stars but also to test theoretical model predictions \citep[e.g.,][]{ribas2000,clar2017,egg2017}. Eclipsing binaries (EBs), in particular, are very powerful tools in astrophysical studies since they allow a direct and accurate determination of fundamental parameters (e.g., masses and radii) of the individual components \citep[e.g.,][]{pie2010,pie2012,torres2010}. These systems have also shown to be very useful in determining precise distances to nearby galaxies \citep[e.g.,][]{bona2006,north2010,vila2010,pie2013,grac2014}, as well as in tracing the structure of the Milky Way \citep[e.g.,][]{hel2013,degri2017}. 

Although several classifications of EBs are currently known, the most frequently used scheme is the one based on the Roche Lobe concept. In this scheme, EBs are classified in three different types depending on the Roche lobe scenario: detached, semi-detached and contact systems. These three types of EBs have been observed using different techniques (astrometry, photometry and spectroscopy) that favour, in particular, the detection of certain types of EBs. The detached EBs, in which their gravitationally bound components are well separated \citep[e.g.,][]{gra2011}, are certainly the most widely studied. The semi-detached EBs, in which one of the components transfers material to the other are less frequently studied, yet they remain numerous \citep{PA2006,PAPA2018A}. Lately, the contact EB systems, in which there is exchange of mass between the two components, have also been well studied. \citet{Jaya2020}, using the All-Sky Automated Survey for Supernovae (ASAS-SN) found a total of 22950 EBs, from which about 43\% are detached Algol-type binaries (EA), 18\% are $\beta$ $Lyrae$ type binaries associated with semi-detached EBs, and almost 39\% are W Ursae Majoris-type binaries (EW) associated with contact EBs.
One of the largest variable star surveys of the inner Milky Way that has been recently completed is the near-infrared (NIR) ESO public Survey VISTA Variables in the V\'ia L\'actea \citep[VVV,][]{min2010,saito2012}. It contains a large amount of still to be extracted photometric data and information about EBs. As a pathfinder project, we describe here our initial efforts towards that direction.

EBs are known to be good distance indicators.  It is then interesting to analyze their period-luminosity (P-L) relations for different EB types (see, e.g., \citealt{AG2015A,chen2016,degri2017} for W UMa systems). In our case, since the luminosity of the EBs cannot be directly obtained, we used parallaxes from the second Gaia data release \citep[see,][]{gaia2016,gaia2018} and the approach described in \citet{luri2018} and \citet{bailer2018} to determine the absolute magnitudes of the EBs.

In this first approach, we present a hundred EB system candidates detected in the NIR images obtained in the VVV Survey, more precisely in the d040 tile, located in the outermost VVV region of the Galactic disc. The Section 2 describes the method we applied to detect variable stars, as well as the procedure chosen to identify and select the EB sample. We describe the methodology used to determine the physical and geometrical parameters of the eclipsing components in Section 3. The analysis and discussion of the results are presented in Section 4, while the main conclusions of our study are summarized in Section 5.

\section{Selected sample}  \label{sec-obs}

\subsection{Observations and data reduction}

The observational data are part of the NIR VVV Survey \citep{min2010,saito2012,hem2014} carried out at the 4.1\,m VISTA telescope at ESO Paranal Observatory (Chile). The VIRCAM camera used for this survey has a 16 NIR detectors array, with a pixel size of 0.34'' \citep{dal2016}. Typically, 70 epochs of observations were acquired in the $K_s$-band between 2010 and 2015. The images were reduced and astrometized by CASU (Cambridge Astronomy Survey Unit) with the VIRCAM pipeline v1.3 \citep{ir2004,eme2004}. More details on the data reduction can be found in \citet{saito2012}. 
Aperture photometry on the individual processed images was also performed by CASU and provided by the VISTA Data Flow System (VDFS). With the obtained photometry, a massive and homogeneous time-series analysis was performed for the point sources detected in the $K_s$-band of the VVV images, plus complementary single epoch observations in the $ZYJH$ bands. Such deep photometry, reaching J and $K_s$ limiting magnitudes of 18.5 and 20.6 respectively, and the multi-epoch $K_s$ band images allowed us to unveil faint variable sources deep into the disc regions of our Galaxy. With this photometry, light curves were generated and analyzed later for variability \citep[see,][]{deka2013,alo2015}.

\begin{table*}
	\small
	\centering
	\caption{Excerpt of basic parameters and solutions derived from the variability analysis for a hundred EB system candidates. The complete version of this table can been seen in the on-line version of this paper.} 
\label{tab1}
\begin{tabular}{ccccccccccr}
		\hline
Source & RA$_{2000}$ & DEC$_{2000}$ & $<K_s>$ & $J-K_s$ & $H-K_s$ & P & Amp. & Parallax* & Parallax &  $d$** \\
EBD040 & (h:m:s) & ($^{\circ}:':''$) & (mag)  & (mag)  & (mag)  & (days) & (mag) &  &  error* &  (kpc)\\
\hline
001	&	11:51:31.50	&	-63:14:10.90	&	13.8	&	0.92	&	0.50	&	1.5753	&	0.27 & 0.12$''$ &	0.09$''$ & 5.078 \\
002	&	11:51:38.90	&	-63:05:08.27	&	14.9	&	0.40	&	0.37	&	0.4358	&	0.51 & . . . & . . . &	. . .\\
003	&	11:51:48.20	&	-62:54:14.80	&	15.0	&	1.50	&	0.61	&	1.4289	&	0.45 & . . . & . . . &	. . .\\
004	&	11:52:23.38	&	-63:15:01.87	&	14.4	&	0.32	&	0.08	&	0.7294	&	0.69 & . . .& . . .&	. . .\\
005	&	11:52:30.00	&	-62:52:05.10	&	14.6	&	0.33	&	-0.09	&	0.5481	&	0.63 & . . .& . . .&	. . .\\
006	&	11:52:54.30	&	-62:30:41.54	&	13.2	&	0.56	&	0.09	&	2.6543	&	0.47 & . . .& . . .&	. . .\\
007	&	11:53:02.90	&	-62:23:41.69	&	14.2	&	1.01	&	0.27	&	4.3012	&	0.80 & . . .& . . .&	. . .\\
008	&	11:53:09.60	&	-62:22:28.52	&	15.3	&	0.78	&	0.23	&	1.0835	&	1.00 &0.08$''$  & 0.20$''$ &4.134 \\
009	&	11:53:10.20	&	-62:12:22.10	&	13.5	&	0.33	&	0.06	&	0.7208	&	0.38 & . . .& . . .&	. . .\\
010	&	11:53:19.54	&	-62:42:24.01	&	15.0	&	0.79	&	0.30	&	1.7052	&	0.61 & . . .& . . .&	. . .\\
\hline
\end{tabular}

\medskip
\tabnote{*Parallaxes and their errors from Gaia-DR2. ** Distance values from \citet{bailer2018}.}

\end{table*}

\begin{figure}
\includegraphics[width=.45\textwidth]{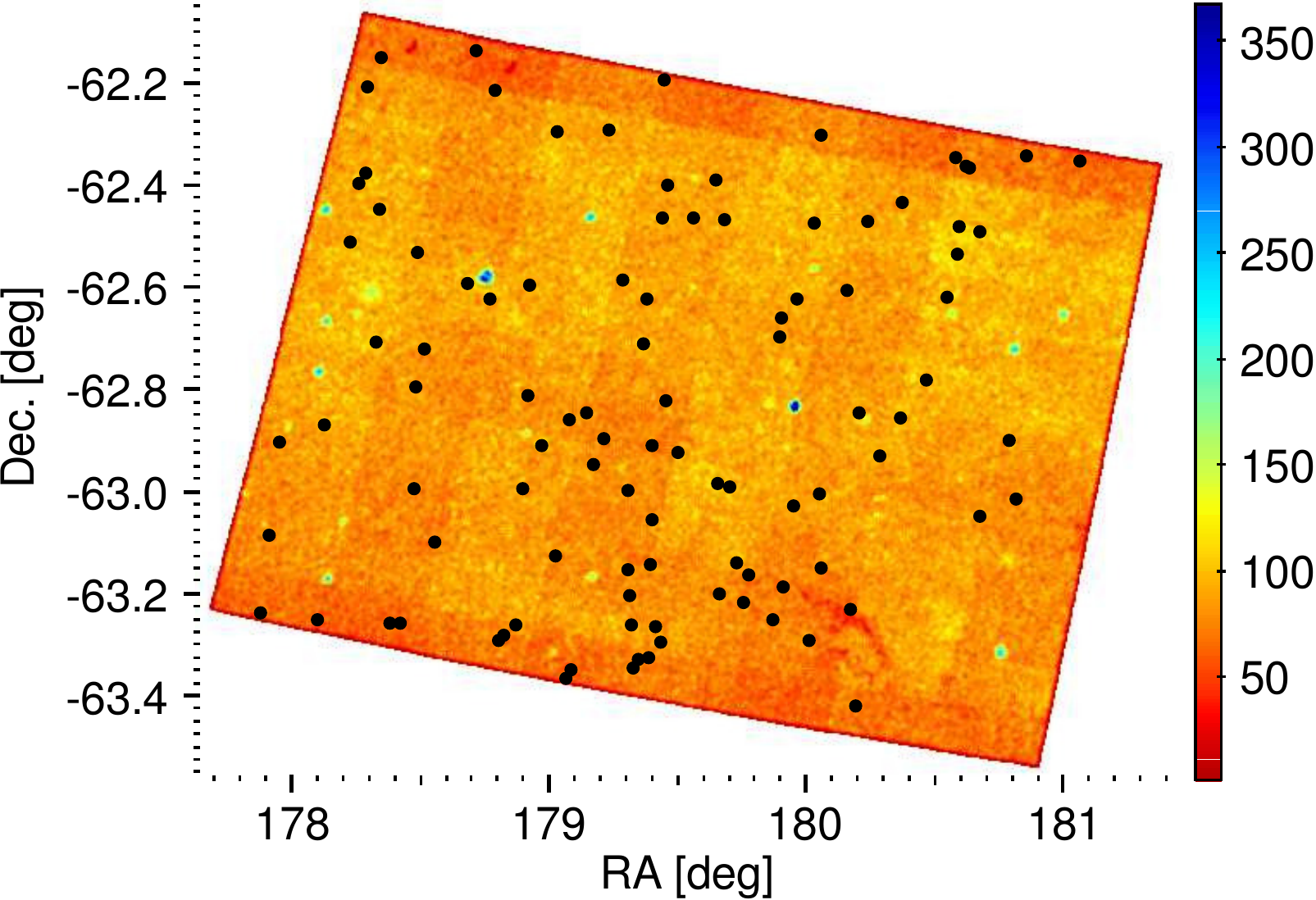}
\includegraphics[width=.45\textwidth]{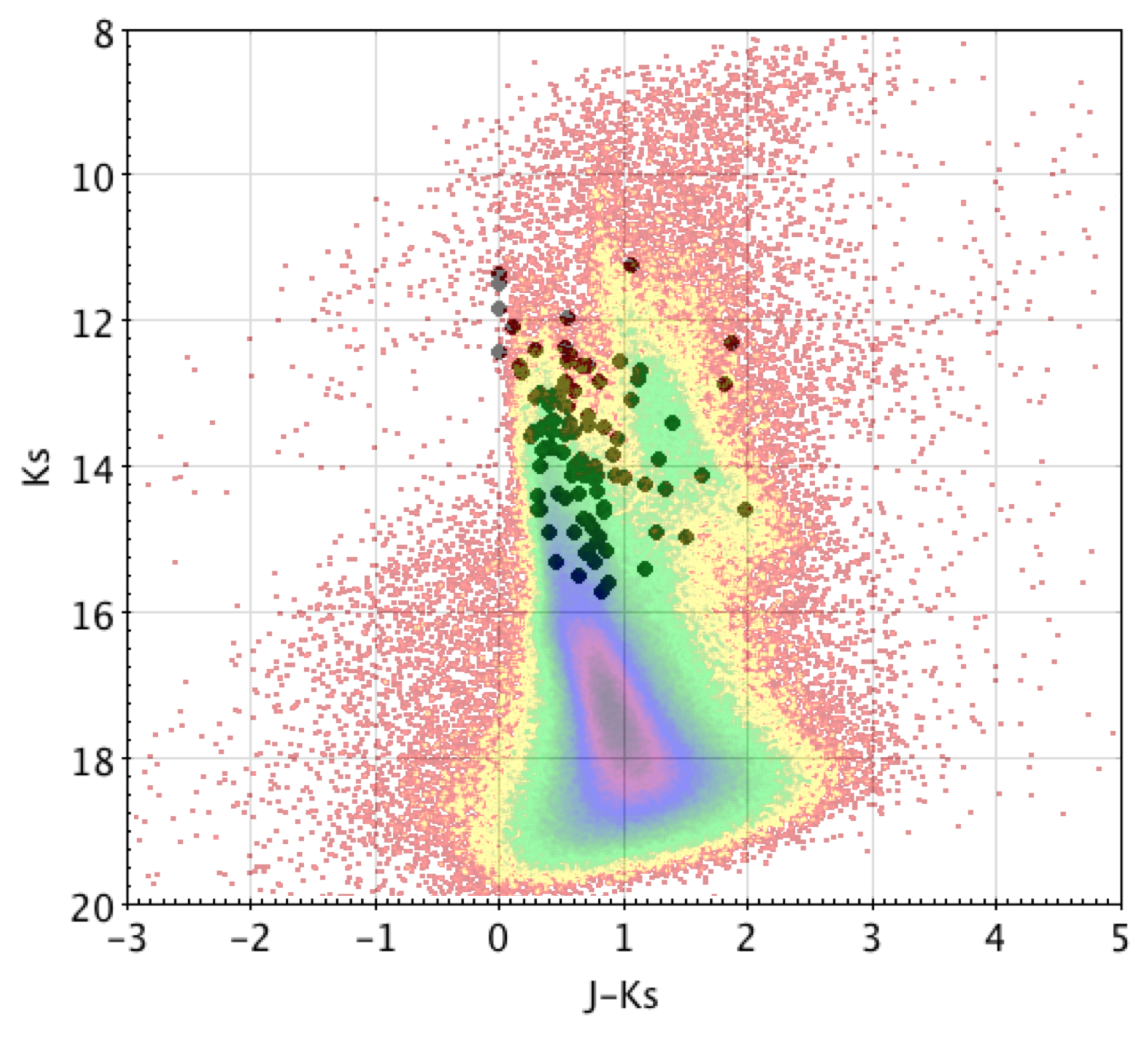}
\caption{Upper panel: finding chart of VVV tile d040. Black filled circles represent the EB systems studied in this work, while the region below indicates the tile colored by its star density.
Lower panel: NIR CMD from the deep PSF photometry built with a procedure similar to \citet{surot19}. The EBs are depicted with black filled circles, while the Hess diagram for the underlying population of the whole d040 tile is shown in colours. There is a clear predominance of disc main sequence stars, with a population of bright and red giants also visible in the upper right portion of the diagram.\label{fk-jk}}
\end{figure}

\subsection{Search for variable stars}

As the first pilot tile to perform the search and analysis of binary system candidates, we selected tile d040 (upper panel of Fig.\ref{fk-jk}) in the outermost Galactic disc region of the VVV Survey. This tile is centered at RA$_{2000}$ = $11^h 58^m 14.16^s$, DEC$_{2000}$ = $-62^{\circ} 48' 15.12''$ (Galactic coordinates l = $296.8962^{\circ}$, b = $-0.5576^{\circ}$),  covering a field of view of 1.64 square degrees \citep[see,][for the tile distribution and nomenclature]{min2010}. The extinction and reddening in tile d040 are comparatively lower than in the surrounding regions (A$_{K_s}$ = 0.62 $\pm$ 0.10 mag, $E$($J-K_s$) = 0.86 $\pm$ 0.14), so tile d040 serves as a NIR window to search for previously "hidden'' objects \citep[see, e.g., ][]{min2018}. Although the procedure for the detection of variables and the  determination of light curves is explained in detail in \citet{alonso2015}, we will briefly introduce it here.

Our first step in the search for variable stars consisted in performing a blind variability exam.  Tile d040 includes $1.6 \times 10^6$ detected sources. Many of those sources with supposed light variations were pre-selected using Stetson variability statistics \citep{ste1996}. As a result of applying this statistics, about 3100 clearly variable stars were detected in the tile d040. As in \citet{tali2016}, we then subjected this first sample of observed variables to a frequency analysis. Two known algorithms were used to perform the signal detection: the Generalized Lomb-Scargle and the Phase Dispersion Minimization (GLS and PDM, respectively, \citealt{ze2009,stel1978}). Once the phase-folded light curves with their preliminary periods were obtained, such light curves were visually examined in order to choose the possible binary system candidates and to eliminate spurious signals. At this stage, about 400 EB candidates were chosen. Then, for this selected sample, we refined the periods iteratively to optimize the light curve fits by using a non-linear Fourier fit, with the first estimated periods obtained from the GLS and PDM analyses. We visually optimized for each source the number of the Fourier order. Finally, we obtained the corresponding light curves for the variable stars, their mean apparent $K_s$ magnitudes and the total amplitudes in the $K_s$-band. For full details of the applied procedure see \citet{AG2015A} and \citet{tali2016}.

\begin{table*}
	\small
	\centering
	\caption{Determined parameters for those EB candidates from our sample included in variable star catalogues.}
\label{tab1b}
\begin{tabular}{ccccc}
		\hline
Source & Other Name & P$_2$* & Variable Class** & References \\
EBD040 &   & (days) & &\\
\hline
5	&	 J115230.02-625205.7	&	NON PERIODIC 	&	ROT &	1\\
	&	GDS\_J1152302-625205	&		&	&2	\\
6	&	 J115254.50-623042.0	&	2.6543714	&	EA & 1	\\
	&	[CKS91] 11504-6213	&		&	Var & 3	\\
9	&	 J115310.22-621222.2	&	NON PERIODIC	&	ROT	&1\\
	&	OGLE-GD-ECL-08166	&		&	& 4	\\
12	&	 J115323.39-620854.3	&	NON PERIODIC	&	ROT	& 1\\
	&	OGLE-GD-ECL-08194	&		&	& 4	\\
13	&	 J115331.49-631526.6	&	1.7237019	&	EA	& 1\\
	&	V0692 Cen*3	&		&	EA	&5\\
16	&	 J115356.45-624739.6	&	2.3598278	&	EA	& 1\\
	&	[CKS91] 11514-6230	&		&	Var	& 3\\
18	&	 J115403.87-624317.6	&	1.1853372	&	EB	& 1\\
	&	GDS\_J1154036-624317	&		&		&2\\
20	&	 J115443.44-623531.6	&	0.6360376	&	EW & 1	\\
	&	[CKS91] 11522-6218	&		&	Var	&3\\
21	&	 J115452.18-620806.6	&	0.5198588	&	EW	& 1\\
	&	OGLE-GD-ECL-08379	&		&	EB	& 4\\
23	&	 J115509.53-621246.8 	&	NON PERIODIC 	&	ROT & 1\\
	&	OGLE-GD-ECL-08416	&		&	& 4	\\
24	&	 J115513.14-631736.3	&	0.4774994	&	EB	& 1\\
29	&	 J115541.98-623545.7	&	1.8492405	&	EA	& 1\\
	&	[CKS91] 11531-6219	&		&	Var & 3	\\
33	&	 J115616.36-632156.3	&	0.8019121	&	EW	& 1\\
	&	GDS\_J1156163-632156	&		&		&2\\
61	&	 J115835.97-622323.4	&	NON PERIODIC/ 0.68 	&	Var & 1	\\
	&	[CKS91] 11560-6206	&		&	Var	& 3\\
70	&	 J115935.48-624149.7	&	1.4025998	&	EA	& 1\\
74	&	 J115951.05-623717.4	&	NON PERIODIC	&	ROT	& 1\\
	&	GDS\_J1159515-623717	&		&		&2\\
77	&	 J120012.73-630011.6	&	1.3822232	&	EB	& 1\\
	&	GDS\_J1200127-630011	&		&		&2\\
83	&	 J120049.26-625039.3	&	0.5270986	&	EB	& 1\\
	&	GDS\_J1200492-625039	&		&		&2\\
92	&	 J120223.37-622855.3	&	193.260715	&	SR	& 1\\
	&	GDS\_J1202233-622855	&		&		&2\\

\hline
\end{tabular}

\medskip
\tabnote{*Period obtained by \citet{ASASc2020} **Variable Class determined by \citet{sos2016} or \citet{ASASc2020}, where are defined EA: Algol-type, EB: $\beta$ $Lyrae$ and EW: W Ursae Majoris-type binaries, Var: Variable star ,ROT: rotating variable star, SR: Semi-regular variable star.\\
References: 1) \citet{ASASc2020}, 2) \citet{GCVS1}, 3) \citet{duque91}, 4) \citet{sos2016}.}

\end{table*}

\subsection{Identification of eclipsing binary candidates}

Visual inspection of the phased light curves enabled us to select a hundred good quality EB candidates. This selection was made by choosing those variables whose light curves exhibit a low or moderate point dispersion and those in which the two minima and a good portion of the maximum can be clearly distinguished. We present in Table \ref{tab1} the selected sample of EB system candidates, together with some parameters associated with each object: equatorial coordinates (J2000), mean  $K_s$-band magnitudes, $J-K_s$ and $H-K_s$ colours, periods in days, $K_s$-band amplitudes, parallaxes in arcsec from the Gaia-DR2 data and the distance $d$ in kpc from \citet{bailer2018}. The last two parameters correspond only to the sources that we found in common with the Gaia-DR2 data. In our sample, we found 55 sources in common with Gaia-DR2 catalogue. However, only 40 of them have well determined parallaxes. Table \ref{tab1} is only partially presented here as guidance in its form and content. The complete version of the table can be found in the on-line version of this article. The lower panel of Fig. \ref{fk-jk} shows the colour-magnitude diagram (CMD) of VVV tile d040, along with the EBs selected for this study. Most of the EB system candidates appear to be located in the region of the main sequence stars, although a handful of them may be evolved red giant stars. As we can see from our sample, none of the detected EBs have mean Ks magnitudes fainter than 16 mag. This is due to the large photometric errors associated to the faintest stars, and thus interfere in the first detection made by Stetson index.\\

Our studied sources were matched with different variable catalogues, such as the ASAS-SN \citep{ASASc2020}, the General Catalogue of Variable Stars (GCVS)  \citep{Samu2017}, the Optical Gravitational Lensing Experiment (OGLE) \citep{sos2016}, the WISE variable \citep{chen2018}, and CDS X-Match Service1\footnote{\url{http://cdsxmatch.u-strasbg.fr/}}. We found a total of 19 stars in our sample that had been previously detected as variable stars (Table \ref{tab1b}). Out of these, 4 were classified as  $\beta$ $Lyrae$ EBs candidates and 8 as $Algol$ and $WUma$ systems. Finally, 7 EBs were classified only as variables or unclassified correctly without a determined period. Table \ref{tab1b} also shows 12 stars classified as binary systems, with periods in agreement with the ones determined in this work (see Table \ref{tab1}).

\section{Determination of physical and geometric parameters}

Once we obtained the observed light curves of the new EB candidates as well as their final amplitudes and periods, we visually classified them as detached, semi-detached or contact binary types, based on the shape of their light curves and on their Roche lobe overflows. Finally, we obtained 50 detached, 13 semi-detached and 37 contact EBs. As expected, the number of contact and detached binary systems is higher than that of semi-detached ones \citep[see, e.g., ][]{PA2006}. 
From the variety of currently available codes for binary system modeling, we opted for the PHysics Of Eclipsing BinariEs \citep[PHOEBE 1.0,][]{prsa2005} code, which is released under the GNU public license. This is a graphical front-end to the Wilson-Devinney code \citep[WD,][]{w1994a,w1994b,w2001,w2006} that has proven to be a very useful tool for EBs analysis. \\

The PHOEBE code fits the light curves in two different steps: a subjective iteration (LC: \textit{Light Curve} process) and an objective iteration (DC: \textit{Differencial Correction} process). In the LC procedure, we can include all known parameters that can be obtained from the theory or observations \citep{w1994a,w1994b,w2001,w2006}, while the DC process is the differential calculus with which the physical and geometric parameters are better determinated. This process permits to estimate the errors associated to parameters obtained from the light-curves modeling. Then, we first analyzed the light curves through the LC procedure using the period inferred from the Fourier analysis. The output of the LC procedure was then used as the input for the DC process. \\

To obtain the best fitting light curve to the observed $K_s$ magnitude values in the LC procedure, we varied other physical parameters like the effective temperatures of the two components ($T_1$ and $T_2$), the mass ratio $q$ ($M_2$/$M_1$), the orbital inclination $i$ of the system, where $i$ = 90$^{\circ}$ means that the observer lies in the plane of the orbit, and the orbital eccentricity value. The relative size of the two components ($R_1$/$R_2$) must also be taken into account. This ratio, however, is not directly modeled but derived from the modeling of the other parameters. In order to model the light curves, it is necessary to assign initial values to the previously mentioned parameters \citep{prsa2005}. \\

After the first light curve model through the LC procedure was obtained, we applied the DC process through which physical parameters are derived. After some iterations, the best fit of the observed light curve was reached. The precision of the fit can be estimated by the $\chi^2$ value, which measures the discrepancy between the observational data and the adopted model. Next, within each iteration, the parameters resulting from each improved fit could be adopted. In every case, there had to be a visual inspection of the obtained results. In case no reasonable agreement was achieved, the corresponding parameters had to be properly changed before the next iteration was made. In some cases, when a wrong EB type was adopted, a final result could not be obtained. Once a possible final solution was achieved, we estimated the uncertainty associated to each of the obtained parameters. 
Although the WD method allows quite a good precision modeling for the parameters, the associated uncertainties increase with the number of variable parameters added in the process of modeling \citep[see, e.g.,][]{prsa2005,nie2017}.\\

\begin{figure}[ht!]
\centering
\includegraphics[width=.48\textwidth]{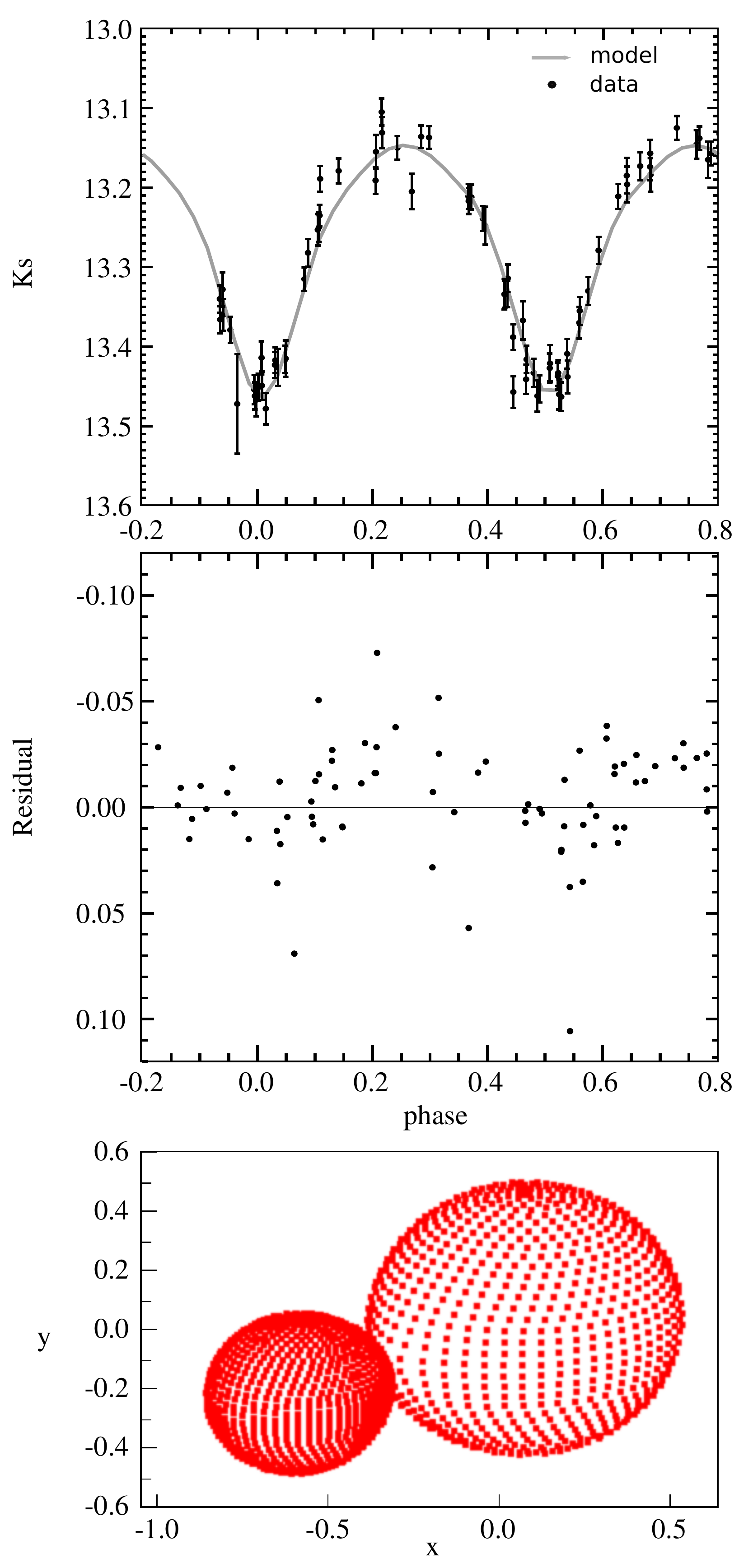}
\caption{Modeling results for the system EBD040-026. The upper panel shows the best modeling fit with the parameters listed in Table \ref{tab2}. The middle panel displays the corresponding error dispersion. The bottom panel recreates the shape that the resulting system should have according to the obtained fundamental parameters.\label{fig1}}
\end{figure}

In order to reach a converging solution, the input parameters should have a value fairly close to the final ones, i.e., the iteration process must be started using a reasonable value of the parameter to be modeled. To do so, we take as initial values of the period, the colour, and the mean $K_s$ magnitudes, those obtained from the analysis described in Section 2.3 (see Table \ref{tab1}). With these values, we build the NIR CMD of tile d040 (lower panel of Fig. \ref{fk-jk}), from which it is possible to estimate an approximate spectral type or the initial value of an effective temperature of the EB candidates. In addition to the mentioned physical parameters, we also modeled the parameters such as the eccentricity and the orbital inclination of the system for which we adopted as initial values 0$^{\circ}$ and 90$^{\circ}$, respectively.
 In general, we find that all types of EB light curves modeling show a high sensitivity to the mass-ratio parameter, for which we had more caution in the modeling process.\\

\begin{figure*}
\hspace{-40pt}
\includegraphics[width=1.\textwidth]{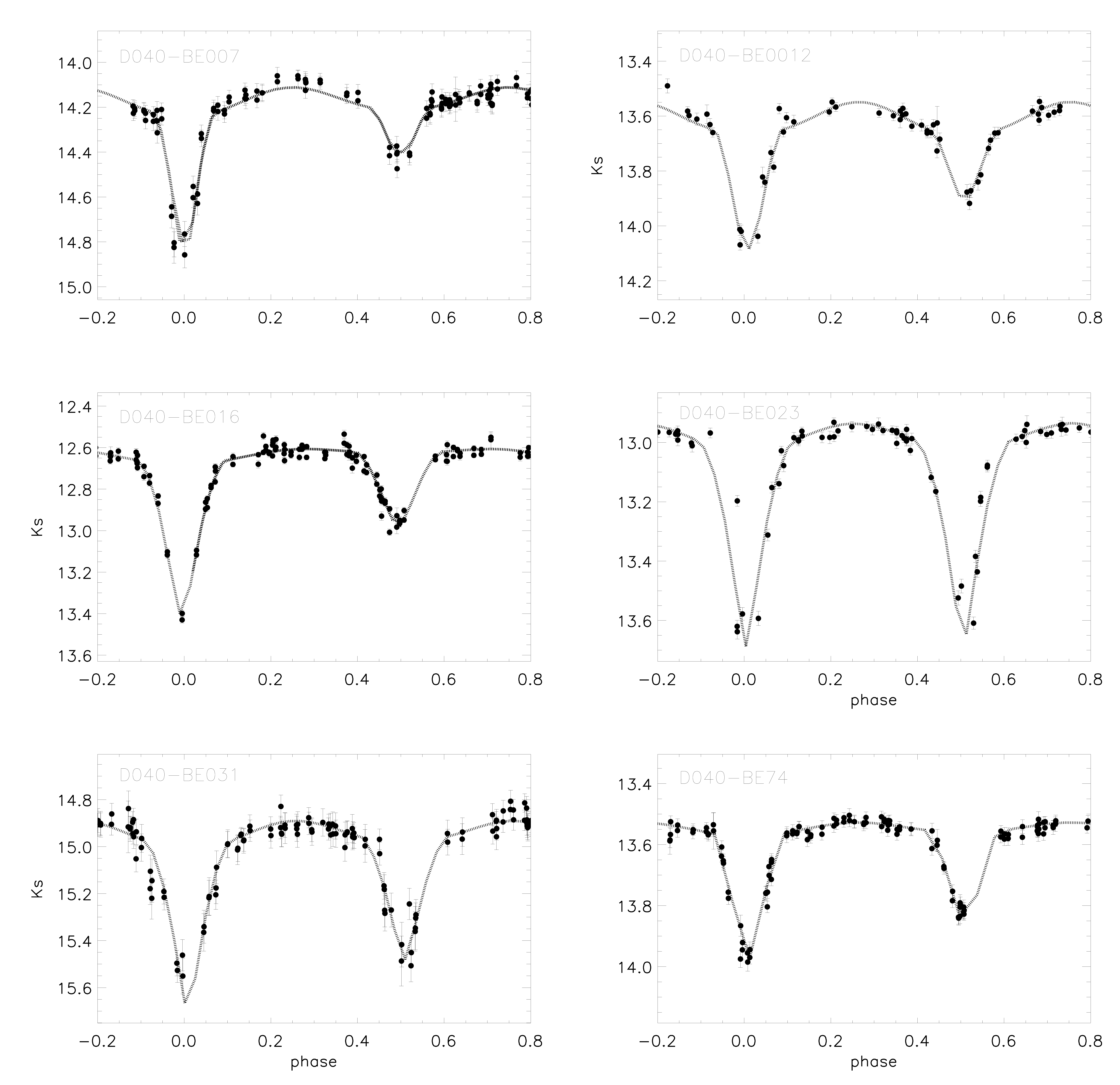}
\caption{Modeled light curves of some of the detached EB candidates.
\label{fig2a}}
\end{figure*}

\begin{figure*}
\hspace{-40pt}
\includegraphics[width=1.\textwidth]{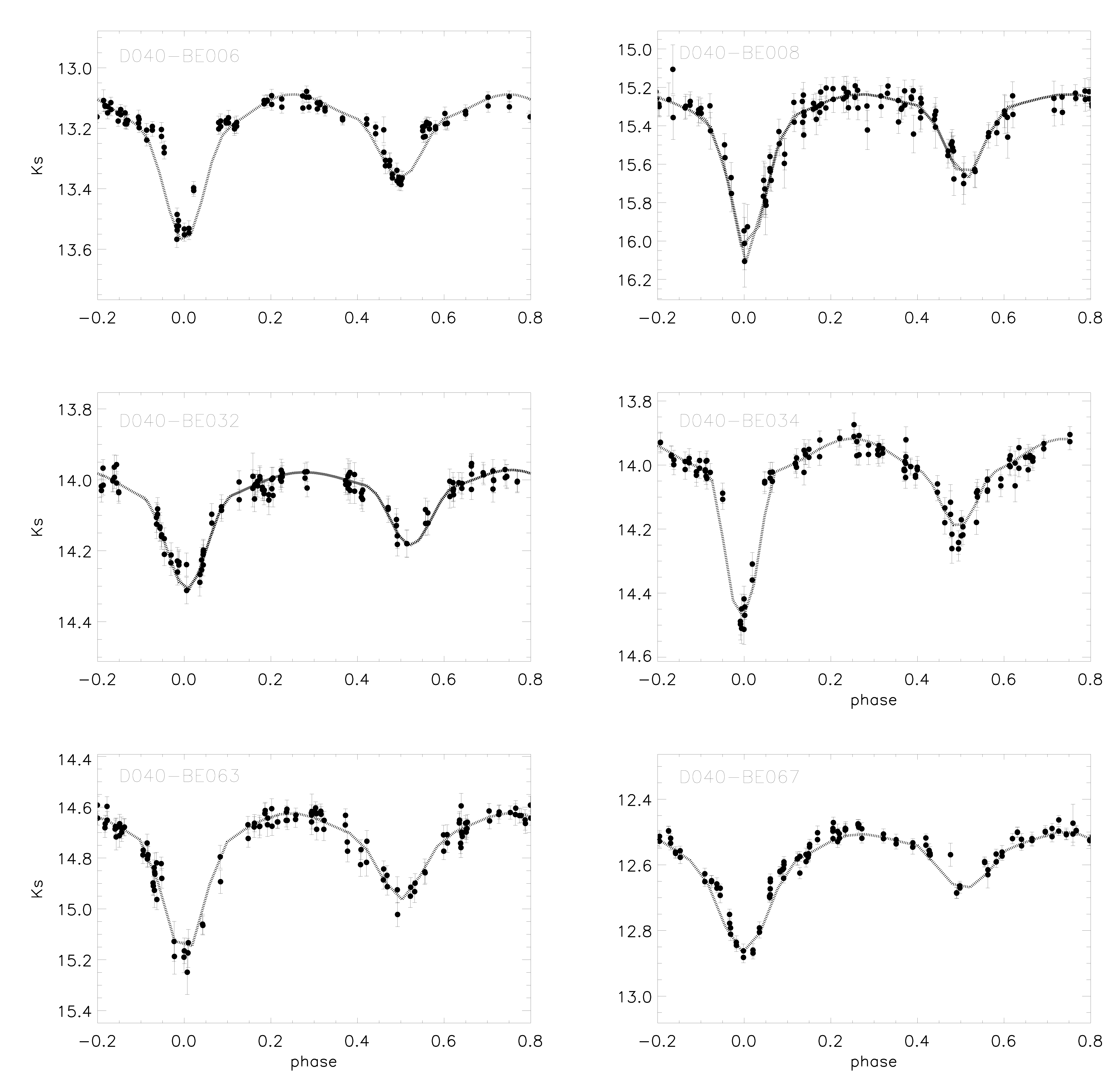}
\caption{Same as Fig. \ref{fig2a} for six semi-detached EB candidates.\label{fig2b}}
\end{figure*}

\begin{figure*}
\hspace{-40pt}
\includegraphics[width=1.\textwidth]{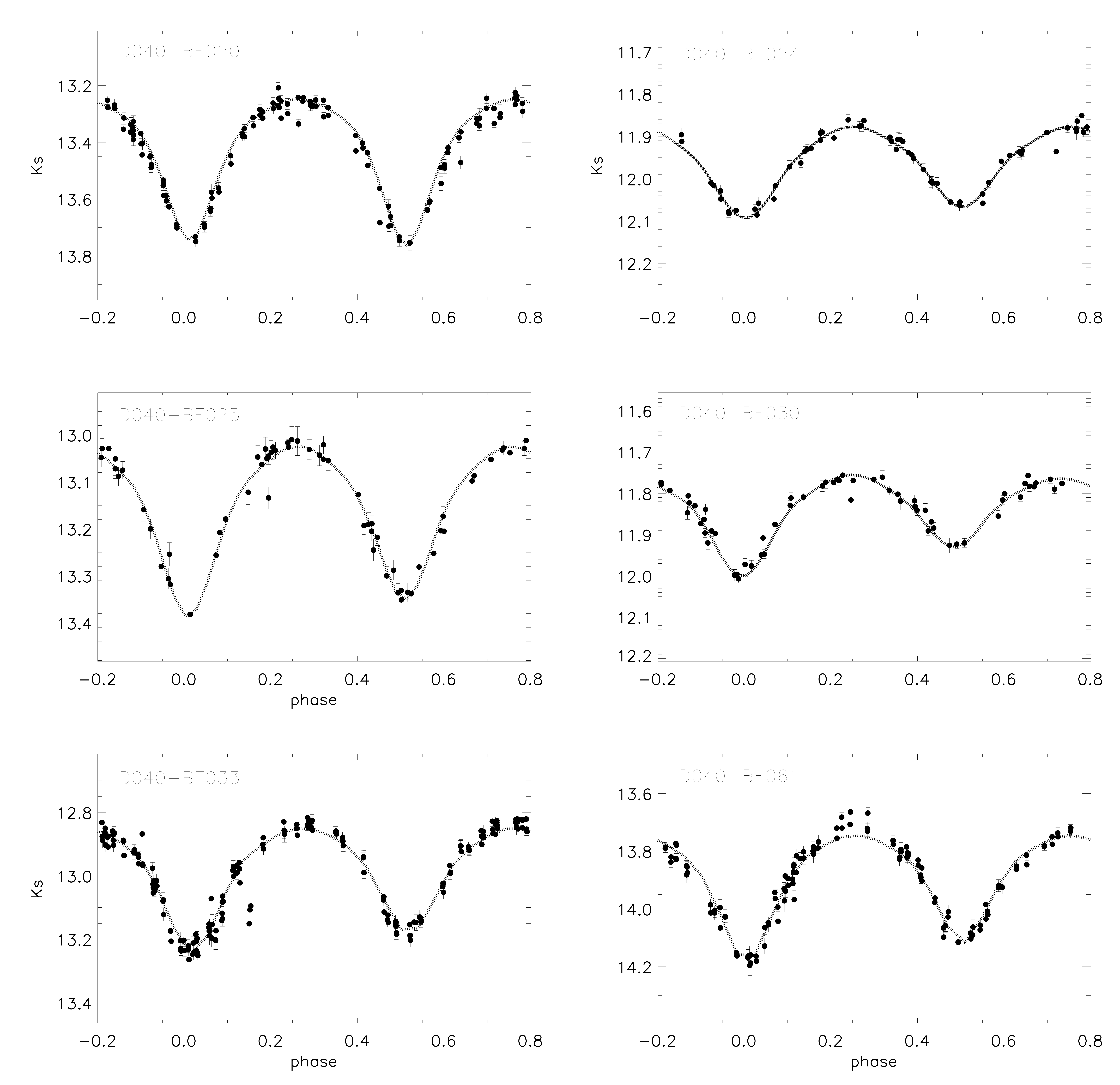}
\caption{Same as Fig. \ref{fig2a} for six contact EB candidates.\label{fig2c}}
\end{figure*}

Summing up, the light curve modeling using PHOEBE was performed by starting with the LC procedure trying to make the theoretical light curve fairly similar to the observed one.
Then, in the following iterations, we visually inspected each fit and its corresponding parameters.
The final solution is reached when $\chi^2$ is both low and stable, while the resulting parameters must have physical meaning. An example of the modeled results for the EBD040-026 system is shown in Fig. \ref{fig1}. This figure exhibits the modeled results and also the dispersion of the errors and shape of the system according to the adopted parameters, respectively. Figures \ref{fig2a}, \ref{fig2b} and \ref{fig2c} show the modeled results for some of the detached, semi-detached and contact EBs, respectively. 
\begin{table*}
	\small
	\centering
	\caption{Excerpt of physical parameters for different types of studied EBs.} \label{tab2}
	\begin{tabular}{ccccccrccc}
		\hline
		Source & Binary class* & $T_1\,$ (K) &  $T_2\,$ (K) &$q$ & $R_1$/$R_2$ & $i\,$($^{\circ}$) & $e$ & $\chi^2$\\
        EBD040 &  &   &   &   &   &   &   &  \\
		\hline
001	&	C	&	5425	$\pm$	1063	&	5717	$\pm$	1944	&	0.3	$\pm$	0.0	&	1.8	&	63.9	$\pm$	1.6	&	0.000	$\pm$	0.008	&	0.27	\\
002	&	C	&	3769	$\pm$	429	&	3426	$\pm$	342	&	0.6	$\pm$	0.0	&	1.3	&	88.3	$\pm$	3.0	&	0.000	$\pm$	0.006	&	2.55	\\
003	&	SD	&	3872	$\pm$	600	&	4374	$\pm$	111	&	1.2	$\pm$	0.1	&	1.7	&	86.7	$\pm$	1.2	&	0.000	$\pm$	0.003	&	3.58	\\
004	&	D	&	5487	$\pm$	815	&	9824	$\pm$	1780	&	0.3	$\pm$	0.0	&	1.6	&	90.0	$\pm$	5.0	&	0.000	$\pm$	0.004	&	0.75	\\
005	&	D	&	3380	$\pm$	267	&	4100	$\pm$	383	&	0.9	$\pm$	0.0	&	2.0	&	92.4	$\pm$	0.9	&	0.000	$\pm$	0.005	&	1.28	\\
006	&	SD	&	3948	$\pm$	436	&	3007	$\pm$	118	&	0.9	$\pm$	0.0	&	1.7	&	78.5	$\pm$	0.7	&	0.005	$\pm$	0.003	&	0.00	\\
007	&	D	&	4991	$\pm$	449	&	2538	$\pm$	92	&	2.7	$\pm$	0.1	&	1.0	&	86.1	$\pm$	0.9	&	0.000	$\pm$	0.003	&	6.39	\\
008	&	SD	&	5983	$\pm$	2134	&	3698	$\pm$	510	&	1.8	$\pm$	0.0	&	1.1	&	83.1	$\pm$	0.8	&	0.000	$\pm$	0.005	&	0.06	\\
009	&	D	&	6433	$\pm$	858	&	6323	$\pm$	837	&	0.8	$\pm$	0.0	&	1.6	&	78.3	$\pm$	0.3	&	0.000	$\pm$	0.003	&	0.32	\\
010	&	D	&	8500	$\pm$	5388	&	3900	$\pm$	1238	&	2.5	$\pm$	0.2	&	0.7	&	75.5	$\pm$	0.9	&	0.000	$\pm$	0.008	&	1.73	\\

		\hline
	\end{tabular}\\
\medskip
\tabnote{* Eclipsing Binary class= D: detached, SD: semi-detached, C: contact.}
\end{table*}

The values derived from the modeling of ten of the studied systems are presented in Table \ref{tab2}, together with their associated errors. The complete table is available in the on-line version of the journal.

\section{Analysis and discussion of results}

\begin{figure}[hb!]
\centering
\includegraphics[width=0.46\textwidth]{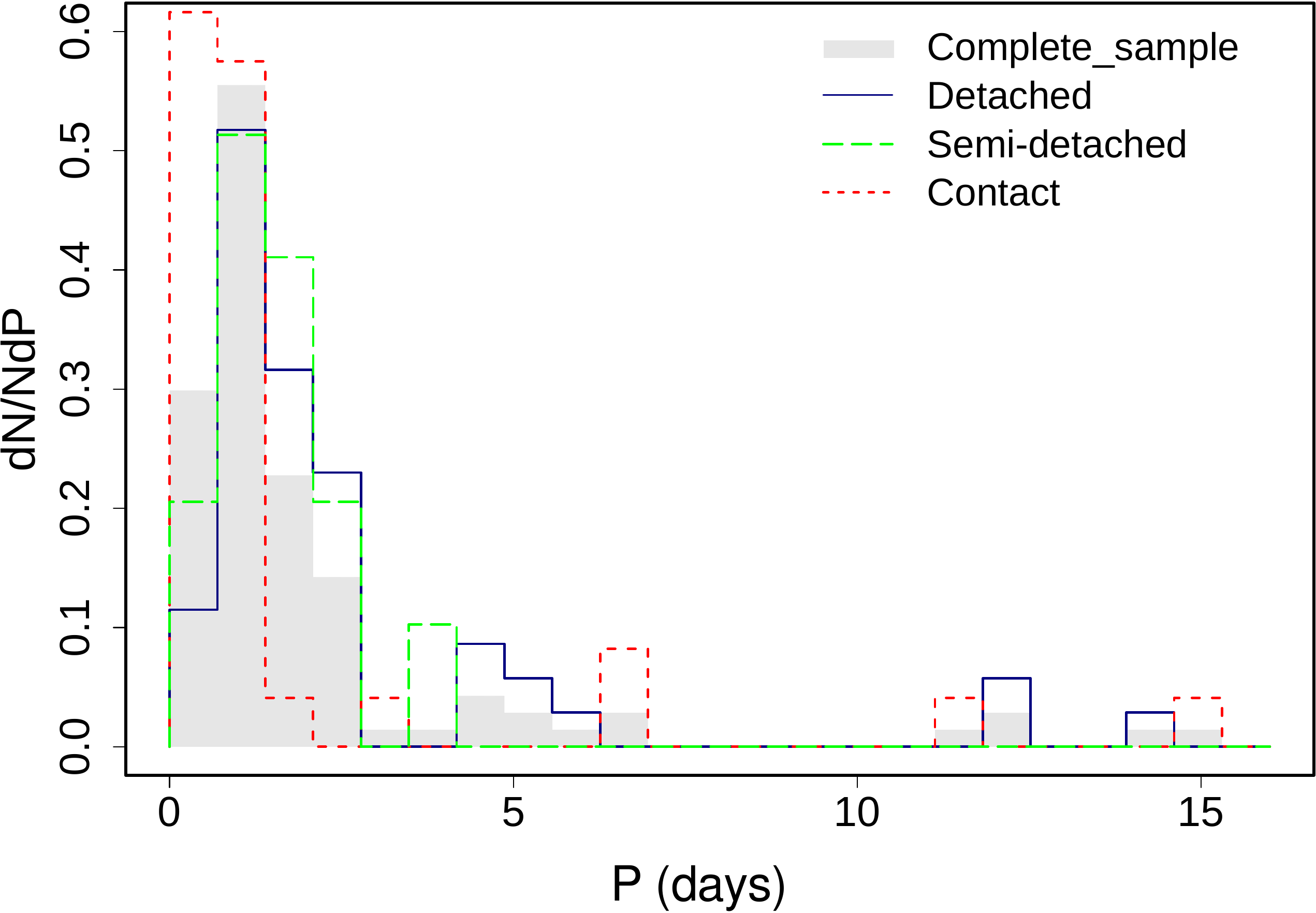}
\includegraphics[width=0.46\textwidth]{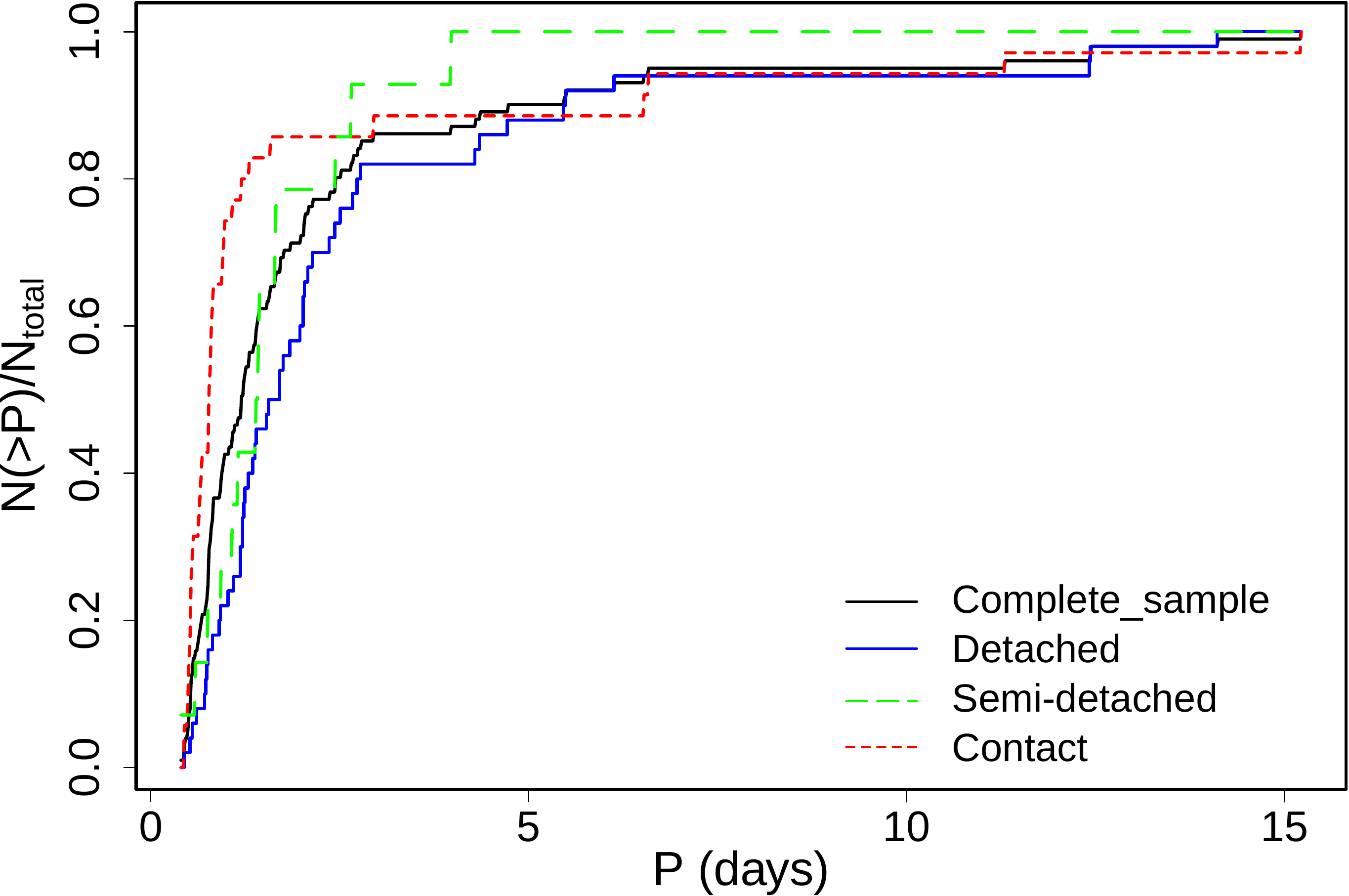}
\caption{The upper panel shows the normalized period histogram of the studied EB sample, while the bottom panel presents its cumulative probability distribution. Subsamples of different types of EBs are represented by dashed lines of different colours: detached EBs in blue, semi-detached EBs in green and contact EBs in red. The total sample is shaded in grey in the upper panel and represented by a grey dashed line in the bottom panel. \label{fig3}}
\end{figure}
We have modeled 100 EBs in the VVV tile d040, obtaining their physical and geometric parameters. This tile has a relatively lower reddening than its surroundings, i.e., we can see deeply into the Galactic plane \citep[see,][]{min2018}. The modeled EBs have mean NIR $K_s$ magnitudes in the range $11.2<K_s<15.6$, with a mean light curve amplitude of 0.80 mag, and relatively short periods in the $0.4 < P $(days) $< 15.2$ range. Fig. \ref{fig3} shows that most of the EBs present periods $P\sim 1.0$ days, the median period being $1.22$ days. 
Our EB distribution of the normalized period histogram, shown in the upper panel of Fig. \ref{fig3}, turned out to be similar to the one obtained by \citet{north2010} for the Small Magellanic Cloud, as well as to the distribution obtained by \citet{PA2006}. \citet{north2010} used radial velocity curves from the VLT and photometric light curves from OGLE, while \citet{PA2006} analyzed a large sample discovered with ASAS. On the other hand, \citet{sos2016} analyzed OGLE data for a sample of $\sim 450 600$ eclipsing and ellipsoidal binaries in the Galactic bulge. They found a period distribution within a range of $0.05 < P $(days) $< 26.00$ peaking at $0.4$ days, while our sample peaks at $P\sim 1.0$ day, with our lower limit ($0.4$ days) coinciding with the \citet{sos2016} peak. Might this result suggest a difference between bulge and disc populations? Since we have only 100 EBs analyzed homogeneously at the moment, we cannot reach a sound conclusion. In further studies, based on a statistically significant EB sample, we hope to address this question properly. We can see in the bottom panel of Fig. \ref{fig3} the corresponding cumulative period distributions of our studied EB sample. Such distribution varies according to the EB type, where in general contact binaries exhibit shorter periods, as expected, with a median value of 0.77 days. This value, however, is larger than one would expect for contact EBs, which may be due to a bias associated either with the tile or with the selection of the sample. Also, we classified two EBs (sources EBD040047 and EBD040062) as contact type that have a period larger than 5 days, with an observational error of $\sim 0.02$ mag in Ks magnitudes and small amplitude ($0.24$ and $0.32$ mag, respectively). In agreement with \citet{ASASc2020}, we suggest that these two sources might be $\beta$ Lyrae type. For the detached and semi-detached EBs, we obtained values for the median period a bit higher (1.71 days and 1.41 days, respectively). \\

On the other hand, the modeling shows that eccentricity values are relatively small and that the orbits are preferentially circular. We found an average value of $\sim 0.004$, within the $0.00 < e < 0.04$ range. For the contact EBs, we derived a median value of the eccentricity of $e\sim 0.0027$, as expected for short period systems (see, e.g., \citealt{chen2018b,ASASc2020}). For the detached and semi-detached EBs, we obtained eccentricities of $e\sim 0.0047$ and $e\sim 0.0058$, respectively. Only two detached EBs, sources EBD040048 and EBD040083, were found to have relatively larger eccentricity values of $e\sim 0.041$ and $e\sim 0.032$, respectively. \\

\begin{figure}
\centering
\includegraphics[width=0.55\textwidth]{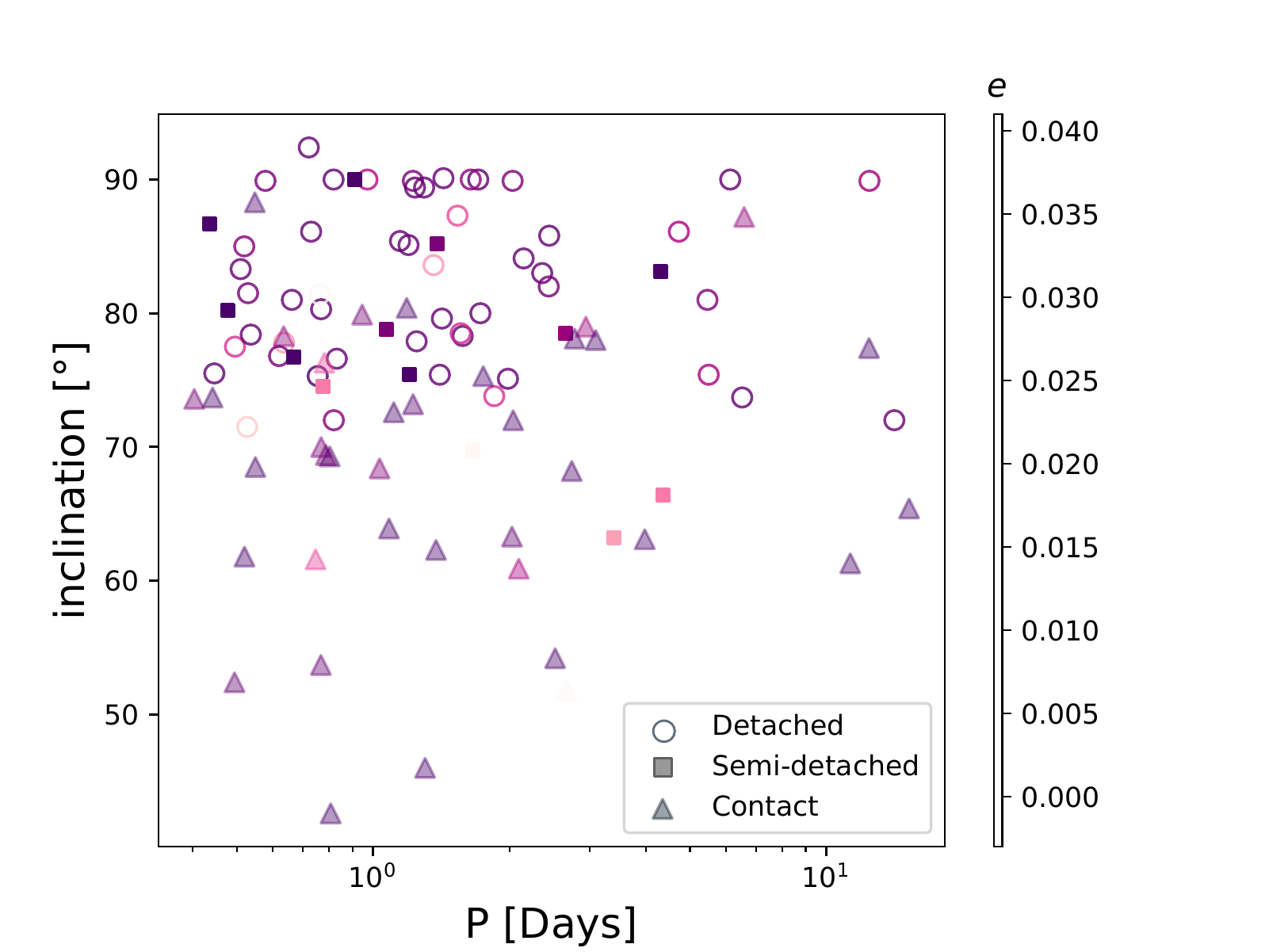}
\caption{Distribution of orbital inclinations as a function of the period for detached (open circles), semi-detached (boxes) and contact (filled triangle) EB candidates. Different colours represent the corresponding eccentricity.  \label{fipe}}
\end{figure}

We analyzed the relation among the parameters $i$, $P$ and $e$. The results can be seen in Fig. \ref{fipe}. Contact EBs with small eccentricities have short to intermediate periods and inclinations lower than $80^{\circ}$. Detached and semi-detached systems present inclinations higher than $70^{\circ}$ and periods distributed along the entire range, although the vast majority of them are shorter than 5 days. Moreover, sources with high eccentricity present inclinations of about $80^{\circ} - 90^{\circ}$ and periods between 1 and 5 days.\\

\begin{figure}
\centering
\includegraphics[width=0.46\textwidth]{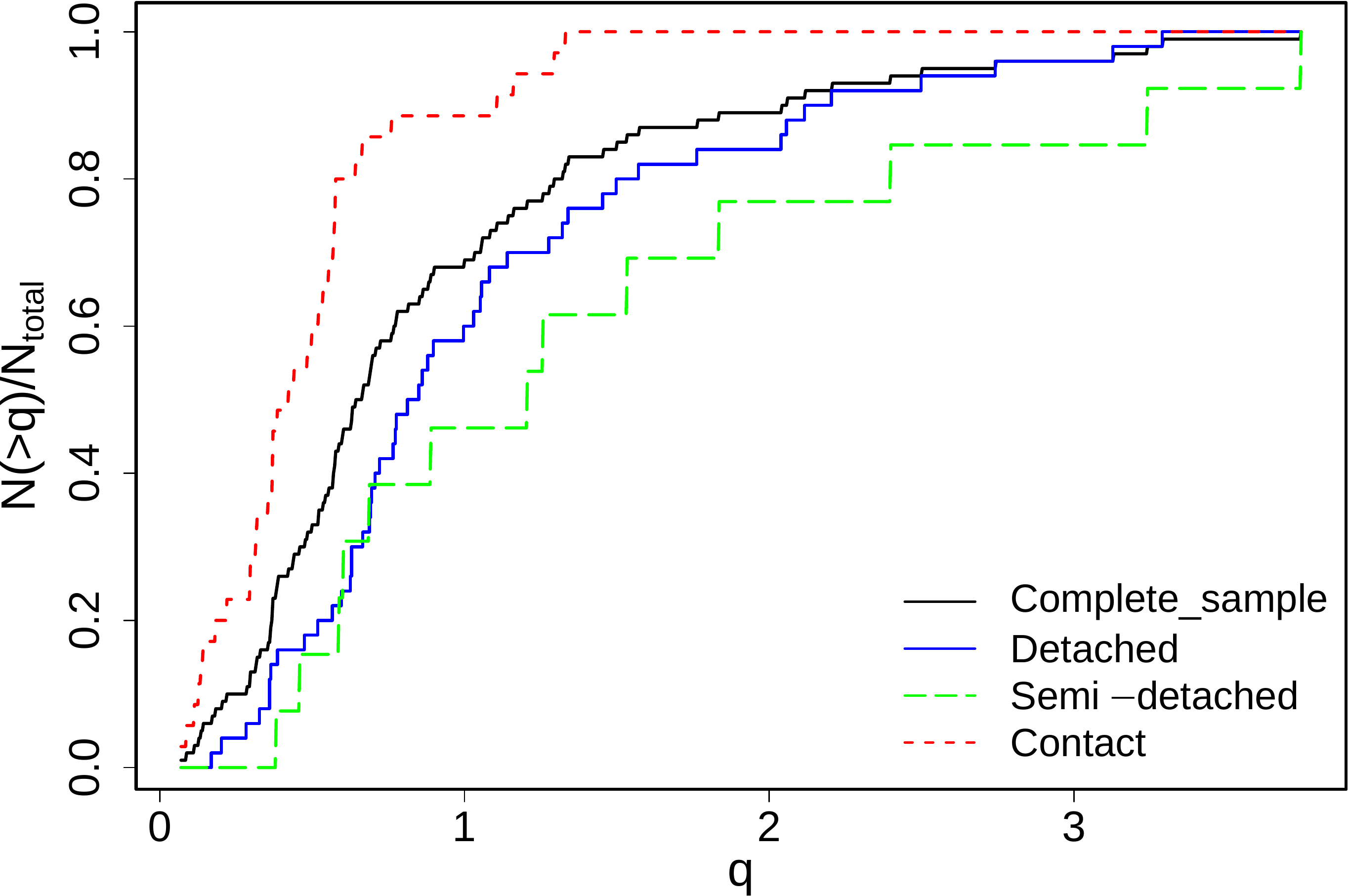}
\caption{The cumulative distribution of the mass-ratio $q$ for the EB candidates. Different colours correspond to different subsamples, i.e., the complete sample (grey), detached EBs (blue), semi-detached EBs (green) and contact EBs (red). The obtained  distributions show that contact EBs have $q$ values lower than 1, while semi-detached binaries have mass-ratios within a wide range of values. These results may depend on which component has filled its Roche lobe.  \label{fq2}}
\end{figure}

As we mention in Section 2, we matched our EB sample to other catalogues of variable stars and confirmed that 19 where previously observed. In general, the classification of these systems in other catalogues is similar to ours. In particular, our periods exhibit good agreement with those derived by \citet{ASASc2020}.
The mass-ratio parameter of our total sample lies within the 0.30 $< q <$ 3.74 range. In order to detect any possible variations in the mass-ratio value for different EB types, we included in Fig. \ref{fq2} the cumulative mass-ratio distributions for our EB candidates. Note in this figure that the contact EBs present $q$ values lower than 1, with a mean $q$ value of $\sim 0.44$. 
However, if we consider the contact and semi-detached systems together, then the mean value of the mass-ratio turns out to be $q$ $\sim$ 0.53, i.e., a little lower than the one derived by \citet{north2010}. On the other hand, the semi-detached EBs present $q$ values higher than 1, a tendency that may be due to a selection effect, i.e., which component of each EB is the mass donnor.

\begin{figure}[ht!]
\centering
\includegraphics[width=0.46\textwidth]{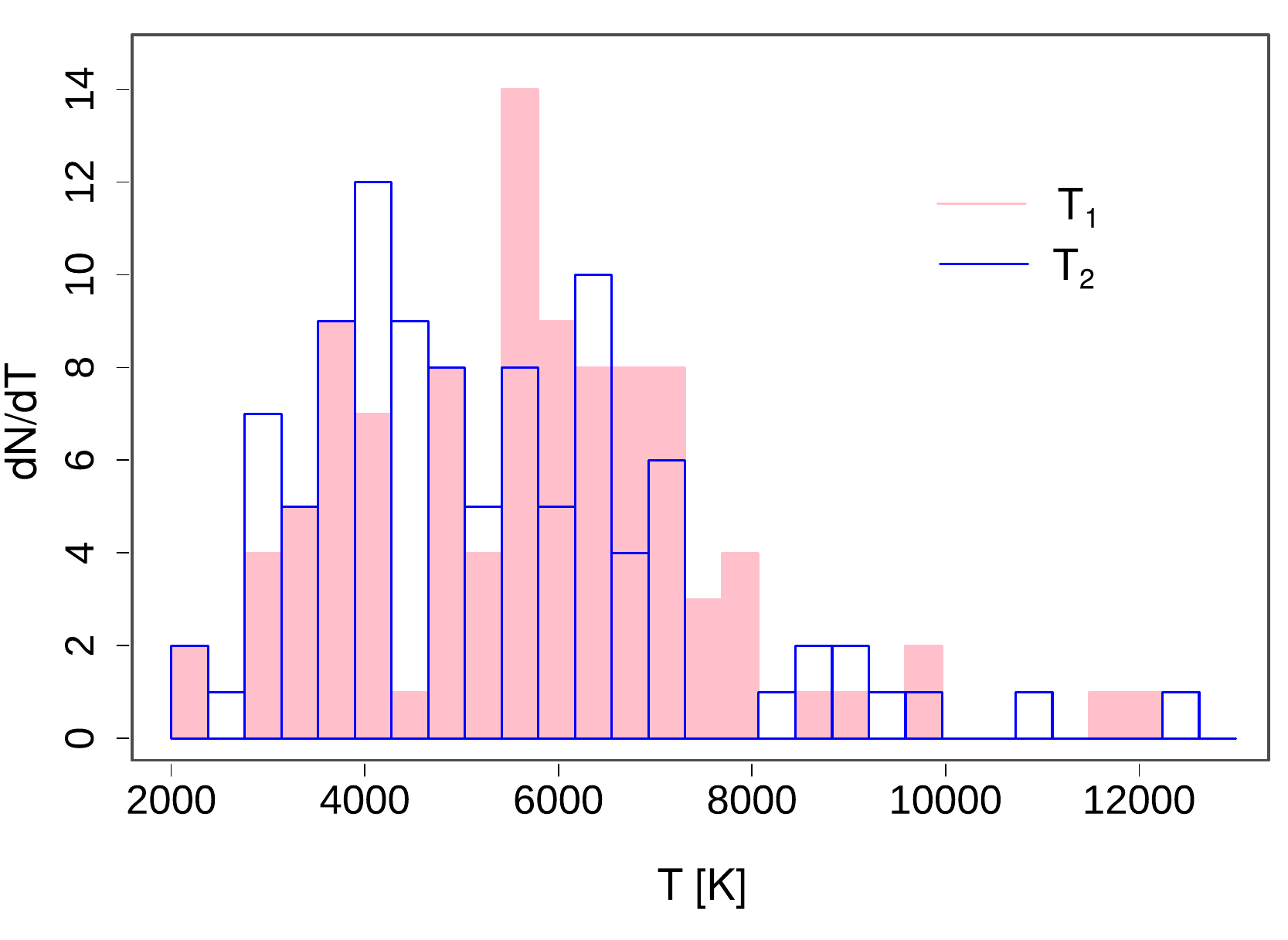}
\includegraphics[width=0.55\textwidth]{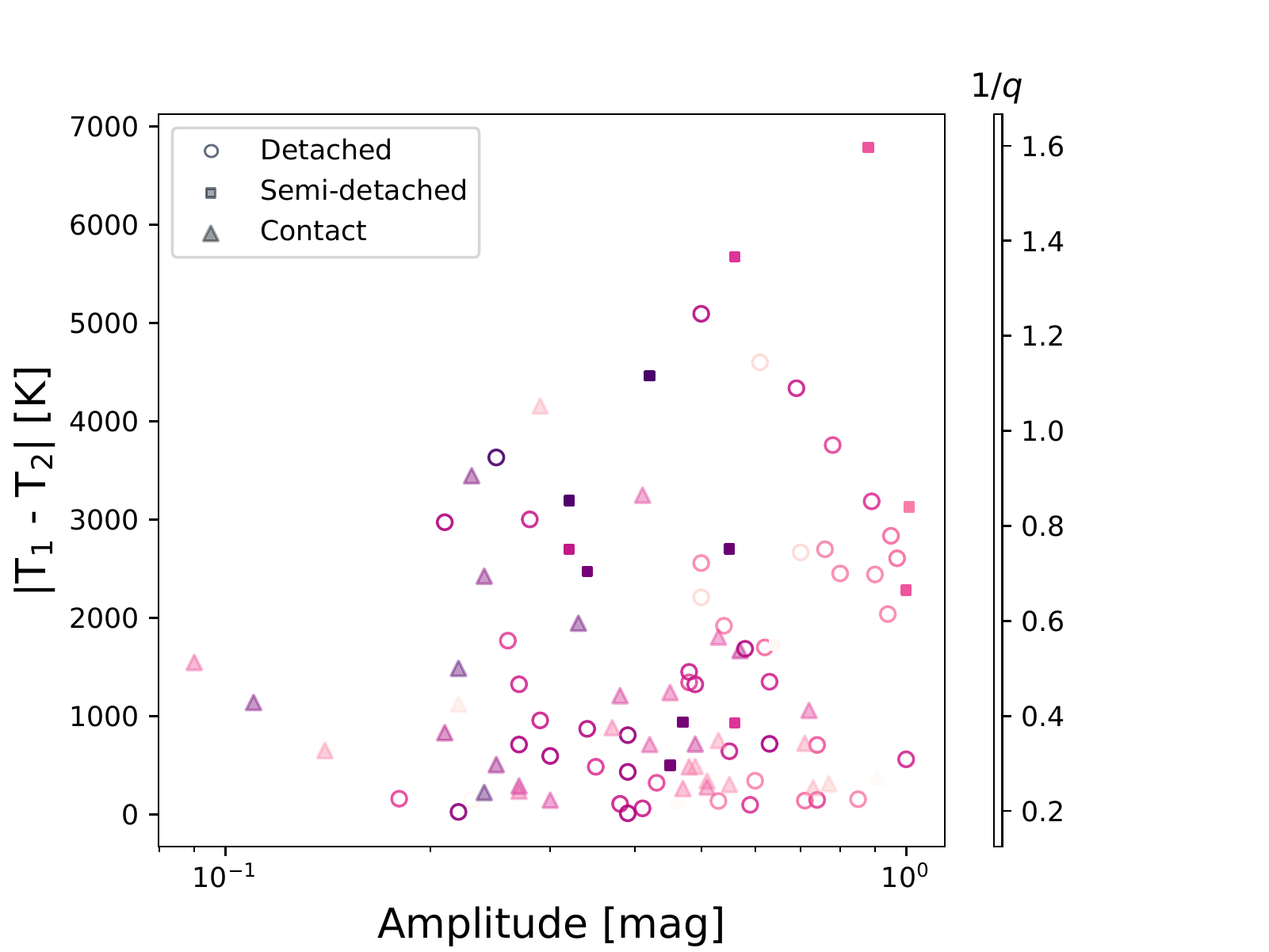}
\caption{Upper panel: Temperature distributions ($T_1$ and $T_2$) derived for the EBs of the studied sample. Lower panel: Distribution of light curve amplitudes as a function of the temperature difference of the components. Symbol colours represent the corresponding EBs mass-ratio values (1/$q$). \label{figt}}
\end{figure}

Other parameters that can be highlighted are the temperatures $T_1$ and $T_2$ of each component, also derived from the PHOEBE model. The upper panel of Fig. \ref{figt} shows the temperature distribution of each component, with peaks at $\sim$ 5800 K and $\sim$ 4000 K for $T_1$ and $T_2$, their mean values being 5700 K and 4900 K, respectively, typical of main sequence stars. The distribution of the temperature difference ( $T_1 - T_2$) derived for the EBs of our sample as a function of the light curve amplitude can be observed in the lower panel of  Fig. \ref{figt}. Different colours indicate their corresponding inverted mass-ratio values (1/$q$), whereas different symbols represent different types of EBs. Though no clear trend is observed in these distributions, the temperature difference in the contact EBs is, on average, smaller than in the detached systems (with a mean value of $\sim$ 700 K and $\sim$ 1300 K, respectively), while their amplitudes are almost all smaller than 1 mag. The sample of semi-detached EBs includes only 13 of them; however, it is noted that the temperature difference between the components covers almost entirely the temperature range with a mean value around $\sim$ 2600 K. These systems present, in general, smaller values for the radii ratio. This means that the semi-detached EBs have almost similar radii even if they may have significant temperature differences in their respective components ($\sim 6000$ $ K$). Such result could be explained if the different types of semi-detached EBs are taken into consideration. One example is the classification made by \citet{malkov2020} in Hot SD, Classical Algols and Cool SD, which also incorporates the possibility of having the inverted components parameters (mass, radius, luminosity) in a sample of 119 semi-detached EBs with the parameters photometrically and spectroscopically determined. However, we have to take into account that the errors associated with the determination of temperatures of our semi-detached sample increase with larger temperature differences.\\

\begin{figure}[hb!]
\centering
\includegraphics[width=0.48\textwidth]{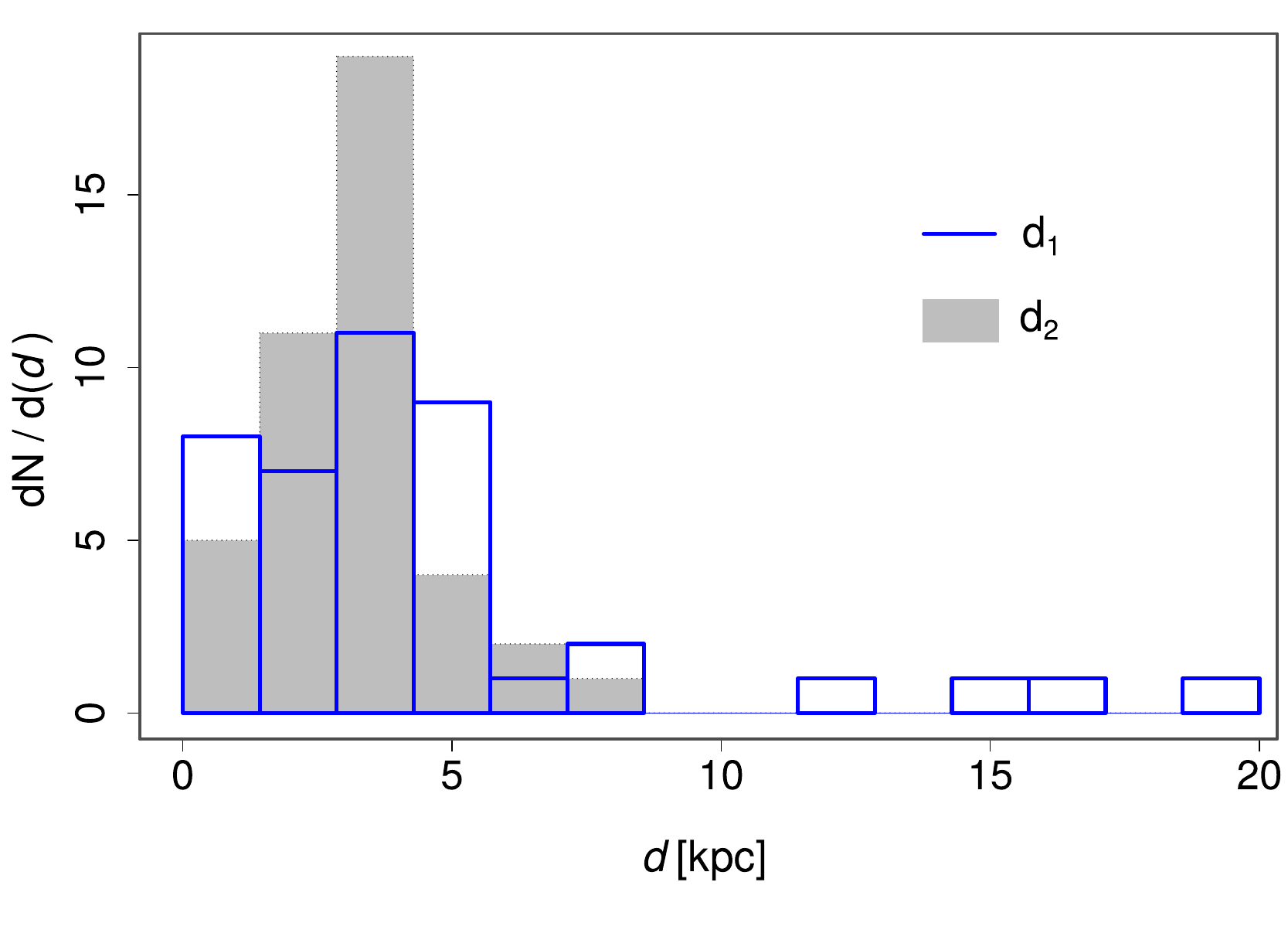}
\caption{Distance histograms for the studied EBs with known parallaxes from Gaia-DR2 (d1) and \citet{bailer2018} (d2). \label{d}}
\end{figure}

Fig. \ref{d} shows the distance distribution measured by Gaia-DR2 \citep{gaia2016,gaia2018}, using the parallaxes listed in Table \ref{tab1}. We notice a wide distribution in distances, while the VVV EBs probe very deep into this region of the Galactic disc. The distances inferred directly by inverting Gaia-DR2 parallaxes (blue-line histogram) have been included in this figure, as well as the Gaia distances corrected by \citet{bailer2018} in the shaded histogram. 
While the two distance distributions are not the same, the difference between them is small. These distances derived from Gaia parallaxes must be seen with caution, since the parallaxes errors some our sources are as large as the parallaxes measurements themselves (see Table \ref{tab1}). We derived $K_s$ absolute magnitudes of each EB from the Gaia-DR2 parallaxes by applying the corresponding corrections from \citet{bailer2018}. These $K_s$ absolute magnitudes were used to analyze the P-L relation including a final sample of 40 sources. Out of these 40 sources, we have only 12 contact EBs for our analysis. A more thorough study of the P-L relation should take into account two possible EB groups: the early-type and the late-type contact EBs. The possibility of a contact binary belonging to one or the other group can be inferred from its location in the Effective Temperature vs. Period (T-P) diagram. Nevertheless, once this T-P diagram was built, many systems of our sample could not be clearly placed in either group. This fact, together with the few EB candidates available for the analysis, did not allow us to reach a clear result for the P-L relationship. It is our aim to expand this study to different regions of the VVV so as to significantly increase the contact EB sample with which carry out a deeper and more reliable study of the P-L relation.\\

\section{Summary and conclusions}

We presented here the light curves and the determination of geometric and physical parameters for a sample of 100 EB system candidates projected on a NIR photometric window in the southern part of the Galactic disc. The EB sample was selected from the VVV database. The comparatively low reddening in the studied region allowed us to obtain precise and deep photometric measurements and light curves in the $K_s$-band. Using the PHOEBE code, we calculated the physical and geometric parameters of the detected EBs. 
The total EB sample is found to be composed of 50$\%$ detached, 13$\%$ semi-detached and  37$\%$ contact binary systems. Their median period is $1.22$ days with an average $K_s$ amplitude of $0.8$ mag. The average period of our EB sample turned out to be higher than the one obtained in the bulge by \citet{sos2016}. Probably, this result may be associated to selection effects, e.g., different wavelengths in which the EBs were observed in the VVV and in the OGLE surveys, respectively. We derived the temperature distributions of each component, which peak at 5800 K and 4000 K for $T_1$ and $T_2$, respectively, with mean values of 5700 K and 4900 K, typical of main sequence stars. In particular, we observed that the differences between the temperature components ($T_1$ - $T_2$) are, on average, smaller in the contact EBs than in the detached systems, while almost all of their respective Ks amplitudes are smaller than 1.0. The cross-matching with Gaia-DR2 data yielded a sample of 40 EBs and only 12 contact systems, so we hope to have a larger contact EB sample to perform a statistically significant study. This new Galactic disc sample is a first approach to the massive study of NIR EB systems. Larger samples of EB system candidates are currently under study, which could lead to a larger statistical EB sample. The method applied here for VVV tile d040 can indeed be used to analyze the remaining tiles of the VVV NIR Survey (196 tiles in the Galactic bulge and 152 in the disc), in which we expect to measure the parameters of thousands of still unknown EB stars. In addition, the future release of Gaia optical light curves for the objects in common with our sample would help to improve the accuracy of their parameters.\\

\begin{acknowledgements}
We gratefully acknowledge data from the ESO Public Survey program ID 179.B-2002 taken with the VISTA telescope, and products from the Cambridge Astronomical Survey Unit (CASU). D.M. acknowledges support from the  BASAL Center for Astrophysics and Associated Technologies (CATA) through grant AFB 170002, and from FONDECYT Regular grant No. 1170121. L.V.G., T.P. and J.J.C. are gratefully indebted to the Argentinian institutions CONICET and SECYT (Universidad Nacional de C\'ordoba) for their support to carry out this research. R.K.S. acknowledges support from CNPq/Brazil through project 305902/2019-9. We thank the referee for the valuable comments and suggestions, which helped to improve the manuscript.

\end{acknowledgements}

\bibliographystyle{pasa-mnras}
\bibliography{ebsVVV}

\end{document}